\documentclass[aps,prb,twocolumn,showpacs,preprintnumbers,floatfix,superscriptaddress]{revtex4}

\usepackage{graphicx}
\usepackage{dcolumn}
\usepackage{subfigure}
\usepackage{wrapfig}
\usepackage{cancel}
\usepackage{color}
\usepackage{textcomp}
\usepackage{amsmath}
\usepackage{amsfonts}
\usepackage{amssymb}
\usepackage{array}

\newcommand{\abs}[1]{\lvert #1 \rvert}

\newcommand{\mean}[1]{\langle #1 \rangle}
\newcommand{\ket}[1]{\lvert #1 \rangle}
\newcommand{\bra}[1]{\langle #1 \rvert}
\newcommand{\expval}[3]{\bra{#1}\hat{#2}\ket{#3}}
\newcommand{\expvalnh}[3]{\bra{#1}#2\ket{#3}}

\begin{document}

\title{Energy density matrix formalism for interacting quantum systems: a quantum Monte Carlo study}
\author{Jaron T.~Krogel}
\affiliation{Materials Science and Technology Division,
Oak Ridge National Laboratory, Oak Ridge, TN 37831, USA}
\author{Jeongnim Kim}
\affiliation{Materials Science and Technology Division,
Oak Ridge National Laboratory, Oak Ridge, TN 37831, USA}
\author{Fernando A.~Reboredo}
\affiliation{Materials Science and Technology Division,
Oak Ridge National Laboratory, Oak Ridge, TN 37831, USA}

\date{\today }

\begin{abstract}
We develop an energy density matrix that parallels the one-body reduced density 
matrix (1RDM) for many-body quantum systems. Just as the density matrix gives 
access to the number density and occupation numbers, the energy density matrix 
yields the energy density and orbital occupation energies. The eigenvectors of 
the matrix provide a natural orbital partitioning of the energy density while 
the eigenvalues comprise a single particle energy spectrum obeying a total 
energy sum rule. For mean-field systems the energy density matrix recovers the 
exact spectrum.  When correlation becomes important, the 
occupation energies resemble quasiparticle energies in some respects.  We 
explore the occupation energy spectrum for the finite 3D homogeneous electron 
gas in the metallic regime and an isolated oxygen atom with ground state 
quantum Monte Carlo techniques implemented in the QMCPACK simulation code.  The 
occupation energy spectrum for the homogeneous electron gas can be described by 
an effective mass below the Fermi level.  Above the Fermi level evanescent 
behavior in the occupation energies is observed in similar fashion to the 
occupation numbers of the 1RDM.  A direct comparison with total energy 
differences shows a quantitative connection between the occupation 
energies and electron addition and removal energies for the electron gas.  For 
the oxygen atom, the association between the ground state occupation energies 
and particle addition and removal energies becomes only qualitative.  The 
energy density matrix provides a new avenue for describing energetics with 
quantum Monte Carlo methods which have traditionally been limited to total 
energies.
\end{abstract}

\maketitle

The single particle or mean field picture has been widely used to explain
the physics of quantum-mechanical systems.  Both qualitative and quantitative 
models based on the notion that individual electrons occupy distinct energy 
levels are indispensible in the analysis of bonding, transport, and optical 
phenomena, among many others.\cite{martin04}  Yet if interactions are fully taken into account, 
the picture becomes more complicated with correlation tangling together the 
previously independent states into a single many-body state.  Despite the added
complexity, it is known that some single particle features are preserved in 
the presence of correlation.  From the success of Fermi liquid theory,\cite{pines66} 
for example, we know that the low-lying excitations in many-body systems 
with weak effective interactions can behave like collections of nearly 
independent quasiparticles.  Even far from the Fermi level, any 
many-body quantum state contains a strong analogy to the single particle 
picture: the natural orbitals and occupation numbers that are made accessible 
from the one body reduced density matrix (1RDM).\cite{lowdin55}  Whether a complementary 
representation of energy levels exists for these single particle states is less 
clear.

Quantum Monte Carlo\cite{foulkes01} methods that deal directly with the 
complexity of the many body problem, usually provide energetic information 
at the level of a single number: the total energy.  While this is 
useful, a more detailed picture of energetics could broaden the interpretive 
power of such methods.  Recent work\cite{krogel13} has shown that the total 
energy of a many-body quantum system can be partitioned across space into an 
energy density.  This hints that an underlying orbital representation of 
energetics might be accessible through the energy density.

In this work we expand the concept of energy density to include non-locality, 
arriving at an energy density matrix.  The one body reduced energy density 
matrix (1REDM) provides energetic information complementary to the established 
1RDM.  In particular, diagonalization of the 1REDM provides a set of natural 
energy orbitals and single particle occupation energies.  Just as in the mean
field picture, the combination of these single particle energies reduces to 
the total energy of the system.  Analogous to the 1RDM, the spectrum of the 
energy density matrix represents a compact partitioning of the total energy 
(rather than the particle number) across states, consisting of strongly occupied 
states below the Fermi energy and weakly occupied states above it.  It will be 
shown that the occupation energies share a close relationship with particle 
addition and removal energies.

The remainder of the paper is organized as follows.
In section \ref{sec:derivation} we derive the non-local energy density matrix 
from the energy density following the example of the 1RDM.  The total energy 
sum rule is proved and a Schr\"{o}dinger-like equation is derived for 
the natural energy orbitals and occupation energies of the matrix.  
Section \ref{sec:exploration} is devoted to an empirical investigation of the 
properties of the occupation energy spectrum via ground state continuum quantum 
Monte Carlo techniques. In section \ref{sec:gsheg} the occupation energy 
spectrum is calculated for a series of finite 
homogeneous electron gases (HEG's) at the same density.  The behavior of the 
spectrum under removal or spin flip of a single electron at constant volume 
for a small HEG is analyzed in section \ref{sec:esheg}.  Section \ref{sec:gsox} 
explores the occupation spectrum of the open shell oxygen atom in its ground state.
A summary of our results is contained in section \ref{sec:summary}.
Details regarding the quantum Monte Carlo evaluation of the energy density 
matrix, as implemented in QMCPACK,\cite{kim12} can be found in appendix \ref{app:qmc_eval}.

\section{Derivation and properties of the energy density matrix}
\label{sec:derivation}
The one body reduced energy density matrix is closely related to the 1RDM.  For 
systems with exchange symmetry, the 1RDM, $n_1(r,r')$, is obtained by 
integrating out all particle coordinates but one from the $N$-body density 
matrix, $\rho_N(R,R')$ and normalizing to the number of particles, $N$:
\begin{align}\label{eq:dmsymm}
  n_1(r,r') = N \int dR_1 \rho_N(r,R_1, r',R_1).
\end{align}
Both here and in the formulae that follow, $R=[r_1,...,r_N]$ denotes all 
particle coordinates, while $R_i=[r_1,...,r_{i-1},r_{i+1},...,r_N]$ is the full 
set with particle $i$ removed.  Spin indices have been suppressed for 
simplicity of discussion.  Rewriting Eq. \ref{eq:dmsymm} more symmetrically and 
introducing the partial trace over $N-1$ particles, 
$Tr_{R_i}(*)\equiv\int dR_i\expvalnh{R_i}{*}{R_i}$, we arrive at the more compact matrix form,
\begin{align}
  \hat{n}_1 = \sum_iTr_{R_i}\hat{\rho}_N
\end{align}
Since the $N$ particle density matrix is normalized to one ($Tr\hat{\rho}_N=1$),
the particle sum rule quickly follows:
\begin{align}
  Tr \hat{n}_1 = \sum_iTr\hat{\rho}_N = \sum_i 1 = N.
\end{align}
The density can be obtained either from the diagonal part of the 1RDM or 
from the expectation value of the density operator, $\hat{n}(r)=\sum_i\delta(r-r_i)$:
\begin{align}
  n(r) = Tr(\hat{n}(r)\hat{\rho}_N) = \expvalnh{r}{\hat{n}_1}{r} 
\end{align}
With this connection in mind, we are now prepared to identify the energy 
density matrix from the evaluation of the energy density.

The energy density operator reflects the partitioning of energy among particles.
Here we take a perspective consistent with the notion of a mean-field, 
\textit{i.e.}, that electrons carry an energy reflecting an average of the 
surrounding particles (this differs from the definition 
in reference \onlinecite{krogel13} where an additive partitioning of energy among all 
particle species was sought).  The energy carried by or belonging to electron 
$i$ can be defined as
\begin{align} \label{eq:hi}
  \hat{h}_i = \hat{t}_i+\hat{v}^{ext}(r_i)+\frac{1}{2}\sum_{j\ne i}\hat{v}^{ee}(r_i,r_j),
\end{align}
which is symmetric under particle exchange.  Here 
$\hat{t}_i=-\tfrac{1}{2}\nabla^2_i$ is the kinetic energy of particle $i$, 
$\hat{v}^{ext}(r)$ represents the external potential, including local Coulomb or 
fully non-local pseudopotential ions, and $\hat{v}^{ee}(r_i,r_j)=1/\abs{r_i-r_j}$ is the 
Coulomb potential between electrons.  The energy density operator simply tracks 
the energy located at point $r$ in real space and reduces to the Hamiltonian 
when integrated:
\begin{align}
  \hat{\mathcal{E}}(r) = \sum_i\delta(r-r_i)\hat{h}_i, \\
  \int dr \hat{\mathcal{E}}(r) = \sum_i \hat{h}_i = \hat{H}.
\end{align}
Anticipating that the energy density matrix (1REDM) $\hat{\mathcal{E}}_1$ can be 
identified in a similar fashion to the 1RDM
\begin{align}
  \mathcal{E}(r) = Tr(\hat{\mathcal{E}}(r)\hat{\rho}_N) 
  = \expvalnh{r}{\hat{\mathcal{E}}_1}{r} 
\end{align}
we evaluate the expectation value of the energy density
\begin{align}
  \mathcal{E}(r) &= Tr(\sum_i\delta(r-r_i)\hat{h}_i\hat{\rho}_N) \nonumber\\ 
  &= \int dR \expvalnh{R}{\sum_i\delta(r-r_i)\hat{h}_i\hat{\rho}_N}{R} \nonumber\\
  &= \sum_i\int dR_i \expvalnh{r,R_i}{\hat{h}_i\hat{\rho}_N}{r,R_i} \nonumber\\
  &= \expvalnh{r}{\sum_iTr_{R_i}\hat{h}_i\hat{\rho}_N}{r}.
\end{align}
We immediately see that the 1REDM can be defined as
\begin{align}
  \hat{\mathcal{E}}_1 = \sum_iTr_{R_i}\hat{h}_i\hat{\rho}_N.
\end{align}

\begin{table}
  \addtolength{\tabcolsep}{3mm}
  \begin{tabular}{llr}
    \hline\hline
    quantity               & formula                                                                        & sum rule\\
    \hline
    number density         &  $\begin{aligned} n(r)=\expvalnh{r}{\hat{n}_1}{r}                                           \end{aligned}$  & $\begin{aligned}\int dr n(r) = N                 \end{aligned}$\\[4mm]
    energy density         &  $\begin{aligned} \mathcal{E}(r)=\expvalnh{r}{\hat{\mathcal{E}}_1}{r}                       \end{aligned}$  & $\begin{aligned}\int dr \mathcal{E}(r) = E       \end{aligned}$\\[4mm]
    number in $\phi_\ell$  &  $\begin{aligned} n_{\phi_\ell}=\expvalnh{\phi_\ell}{\hat{n}_1}{\phi_\ell}                     \end{aligned}$  & $\begin{aligned}\sum_\ell n_{\phi_\ell} = N       \end{aligned}$\\[4mm]
    energy in $\phi_\ell$  &  $\begin{aligned} \mathcal{E}_{\phi_\ell}=\expvalnh{\phi_\ell}{\hat{\mathcal{E}}_1}{\phi_\ell} \end{aligned}$  &$\begin{aligned}\sum_\ell\mathcal{E}_{\phi_\ell}=E \end{aligned}$\\[4mm]
    number spectrum        &  $\begin{aligned}  \hat{n}_1\ket{\eta_\ell} = n_\ell\ket{\eta_\ell}                          \end{aligned}$ & $\begin{aligned}\sum_\ell n_\ell = N             \end{aligned}$\\[4mm]
    energy spectrum        &  $\begin{aligned}  \hat{\mathcal{E}}_1\ket{\xi_\ell} = \mathcal{E}_\ell\ket{\xi_\ell}        \end{aligned}$ & $\begin{aligned}\sum_\ell \mathcal{E}_\ell = E   \end{aligned}$\\[4mm]
    \hline\hline
  \end{tabular}
  \caption{Comparison of related quantities derived from the 1-body reduced density 
    $\hat{n}_1$ and energy density $\hat{\mathcal{E}}_1$ matrices, including 
    the density, occupation of an arbitary orbital $\phi_\ell$ (assumed to 
    be part of a complete set), and eigenvalue/occupation spectrum.  $N$ is the total 
    number of particles and $E$ is the total energy.}
  \label{tab:comp}
\end{table}

From the definition of the 1REDM several properties quickly become apparent.
Since many of these properties are directly analogous to those of 
the 1RDM we present a side-by-side comparison in table \ref{tab:comp}.
Perhaps the most interesting of these is that the energy 
density matrix provides access to a single particle energy spectrum embedded 
in the many body state.  Minimizing the energy 
$\mathcal{E}_\xi=\expvalnh{\xi}{\hat{\mathcal{E}}_1}{\xi}$ contained in an 
arbitrary state $\ket{\xi}$ results in a set of natural energy orbitals 
$\ket{\xi_\ell}$ and occupation energies $\mathcal{E}_\ell$
\begin{align}\label{eq:eigval}
  \hat{\mathcal{E}}_1\ket{\xi_\ell} = \mathcal{E}_\ell\ket{\xi_\ell}.
\end{align}
It can be shown that certain quantum numbers describing the many-body state 
(\emph{e.g.} crystal momentum) are transferred to the natural energy orbitals.

For a mean-field system, the occupation energy spectrum is identical to the 
spectrum of the 1-body Hamiltonian.  This can be demonstrated by calculating 
the explicit representation of the energy density matrix for the mean field 
system.  In this case, the $N$-body Hamiltonian is given by the sum of identical 
1-body operators $\hat{H}^{MF} = \sum_i\hat{h}^{MF}(r_i)$ and the wavefunction 
is a Slater determinant\cite{slater29} of the occupied orbitals, denoted $\Psi^{MF}$, with 
the orbitals obeying the single particle Schr\"{o}dinger\cite{schrodinger26} equation 
$\hat{h}^{MF}\ket{\phi_n^{MF}}=\mathcal{E}_n\ket{\phi_n^{MF}}$.  The energy density 
matrix can be obtained in real space following a process similar to the 
derivation of the Slater-Condon\cite{slater29,condon30} rules for one-body operators.  
\begin{align}
  \mathcal{E}_1(r,r') &= \sum_i\int dR_i \hat{h}^{MF}_{r}\Psi^{MF}(r,R_i)\Psi^{MF*}(r',R_i) \nonumber \\
                      &= \sum_i \hat{h}^{MF}(r) \phi^{MF}_i(r)\phi^{MF*}_i(r') \nonumber \\
                      &= \sum_i \mathcal{E}^{MF}_{i} \phi^{MF}_i(r)\phi^{MF*}_i(r') 
\end{align}
This is just the spectral representation of $\hat{h}^{MF}$ and so the 
eigenvalues and eigenvectors of the energy density matrix are $\{\mathcal{E}^{MF}_n\}$ 
and $\{\ket{\phi^{MF}_n}\}$, respectively.  
Since $\{\ket{\phi^{MF}_n}\}$ are also the natural orbitals in this case 
($\ket{\eta^{MF}_n}=\ket{\phi^{MF}_n}$), any deviation between $\{\ket{\eta_n}\}$ 
and $\{\ket{\xi_n}\}$ for an interacting system is a direct indication of the 
strength of correlation effects beyond what can be captured by an effective mean field.
For many systems that are not strongly correlated, the natural orbitals and the natural 
energy orbitals are expected to be similar.  

A characteristic property of a mean-field system is that the total energy is just 
the sum of the single particle energies $E^{MF} = \sum_{i=1}^N\mathcal{E}^{MF}_i$.  This 
property is preserved in the spectrum of the energy density matrix even for interacting 
systems because it obeys the following total energy sum rule
\begin{align}\label{eq:Esum}
  Tr \hat{\mathcal{E}}_1 = Tr\sum_i\hat{h}_i\hat{\rho}_N = Tr(\hat{H}\hat{\rho}_N) = E
\end{align}
which is similar to the particle sum rule of the 1RDM $Tr\hat{n}_1=N$.  The 
spectral representation of the sum rule is similar to the mean field form
\begin{align}
  \sum_{\ell=1}^\infty \mathcal{E}_\ell = E.
\end{align}
It is expected that states with vanishing occupation number will also make 
vanishing contributions to the total energy, provided the occupation 
spectrum is bounded from below and non-oscillatory.

Unlike the mean-field case, the total energy for an interacting system is not 
a linear functional of the occupation numbers.
Factoring the energy density matrix in terms of one- and two-body contributions gives
\begin{align}\label{eq:f12}
  \hat{\mathcal{E}} = (\hat{t}+\hat{v}^{ext})\hat{n}_1 + \hat{v}^{ee}_1.
\end{align}
Here $\hat{t}$ and $\hat{v}^{ext}$ are the one-body kinetic and external potential 
operators given in Eq. \ref{eq:hi}, $\hat{n}_1$ is the 1RDM, and $\hat{v}^{ee}_1$ 
is the one-body reduced electron-electron interaction matrix given by 
\begin{align}
  \hat{v}^{ee}_1 = \sum_iTr_{R_i}\bigg(\frac{1}{2}\sum_{j\ne i}\hat{v}^{ee}(r_i,r_j)\hat{\rho}_N\bigg).
\end{align}
Upon expanding the total energy in the basis of natural orbitals
\begin{align}
  E &= \sum_\ell \expvalnh{\eta_\ell}{\hat{\mathcal{E}}_1}{\eta_\ell} \nonumber\\
    &= \sum_\ell \left[n_\ell \expvalnh{\eta_\ell}{\hat{t}+\hat{v}^{ext}}{\eta_\ell} + \expvalnh{\eta_\ell}{\hat{v}^{ee}_1}{\eta_\ell} \right]
\end{align}
we see that the one body terms contribute linearly in occupation number while 
the two body interaction term involving $\hat{v}_1^{ee}$ will in general contain 
higher order contributions.  This is consistent with expectations based on 
Landau's theory of quantum fluids\cite{landau57} where terms up to quadratic order 
in occupation number are retained in the total energy functional.

The natural energy orbitals obey a Schr\"{o}dinger-like integro-differential equation, 
which can be obtained by combining expressions \ref{eq:eigval} and \ref{eq:f12} and 
projecting into real space:
\begin{align}\label{eq:sch}
  \left[-\frac{1}{2}\nabla^2_r+v^{ext}(r)\right] \int dr' n_1(r,r')&\xi_\ell(r') \nonumber\\
     + \int dr' v^{ee}_1(r,r')&\xi_\ell(r') = \mathcal{E}_\ell\xi_\ell(r)
\end{align}
The form of this equation resembles the quasi-particle equation arising in $GW$ theory.\cite{hedin65}
The resemblance in form becomes stronger in the case of weak correlation ($\xi_\ell(r)\approx\eta_\ell(r)$)
\begin{align}\label{eq:gw}
  \left[-\frac{1}{2}\nabla^2_r+v^{ext}(r)\right] &\xi_\ell(r) \nonumber\\ 
    + \int dr' \frac{v^{ee}_1(r,r')}{n_\ell}&\xi_\ell(r') = \frac{\mathcal{E}_\ell}{n_\ell}\xi_\ell(r)
\end{align}
with $v^{ee}_1(r,r')/n_\ell$ and $\mathcal{E}_\ell/n_\ell$ filling the roles of self-energy and 
quasi-particle energy level, respectively.  It should be stressed at this point that the 
eigenvalues of Eq. \ref{eq:sch} formally correspond to particle removal energies in the 
weakly-interacting limit, in contrast to $GW$.   Another important distinction to keep in mind 
is that Eqs. \ref{eq:sch} and \ref{eq:gw} provide a single particle energy spectrum
measured within a particular many body state, $\hat{\rho}_N$.  The full single particle
spectrum is obtained by combining measurements over all states in the many-body spectrum.

Defining the energy of a particular level as 
\begin{align}
  \bar{\mathcal{E}}_\ell = \frac{\mathcal{E}_\ell}{n_{\xi_\ell}}  = \frac{\expvalnh{\xi_\ell}{\hat{\mathcal{E}}_1}{\xi_\ell}}{\expvalnh{\xi_\ell}{\hat{n}_1}{\xi_\ell}}
\end{align}
is sensible because $\bar{\mathcal{E}}_\ell$ shifts rigidly with a constant shift 
in the external potential while $\mathcal{E}_\ell$ does not (to see this, note from 
Eq. \ref{eq:f12} that $\hat{\mathcal{E}}_1\rightarrow \hat{\mathcal{E}}_1 + v_c\hat{n}_1$ 
when $\hat{v}^{ext}\rightarrow\hat{v}^{ext}+v_c$).    
Since the number of electrons in the level is 
$n_{\xi_\ell}=\expvalnh{\xi_\ell}{\hat{n}_1}{\xi_\ell}$, it is consistent that 
the energy in the level is the product of the energy level and its occupation, or 
$n_{\xi_\ell}\bar{\mathcal{E}}_\ell = \mathcal{E}_\ell$, which is the portion of the 
total energy attributed to $\ket{\xi_\ell}$.  

By distinguishing an energy level from the amount of energy occupied in that level, 
we can further partition the energy 
among spin species where spin interactions are small enough to be neglected.
In this case a well-defined spin state can be formed by assigning each electron 
to have an up or down spin and the 1RDM, along with its corresponding occupation 
numbers, can be factored into up and down components.  The occupation energies 
for spin up or down electrons can then be defined as
\begin{align}\label{eq:Eocc_spin}
  \mathcal{E}^{\uparrow/\downarrow}_\ell = n^{\uparrow/\downarrow}_{\xi_\ell}\bar{\mathcal{E}}_\ell = \frac{n^{\uparrow/\downarrow}_{\xi_\ell}}{n_{\xi_\ell}^\uparrow+n_{\xi_\ell}^\downarrow}\mathcal{E}_\ell.
\end{align}
This relation has been used in all following sections to obtain spin-resolved 
occupation energies.

\section{Exploration of the occupation energy spectrum}
\label{sec:exploration}
In the preceeding analysis both rigorous connections and analogies to mean-field 
behavior have been made, but a more empirical investigation of the remaining properties 
of the occupation energy spectrum is clearly warranted.  In the following two sections 
we analyze the occupation spectrum for the fully interacting 3D homogeneous electron 
gas (HEG), first for the ground state over a range of system sizes and then for 
excited states of a small system.  Study of the excited state systems probes the 
relationship between the occupation energies and single particle addition and removal 
for the HEG.
Following the analysis of the HEG we explore the partitioning of the energy among 
core and valence states for the open shell ground state of the oxygen atom.

\subsection{Ground state occupation energies of the homogeneous electron gas}
\label{sec:gsheg}
The homogeneous electron gas is an ideal test case to explore the properties of the occupation 
energy spectrum.  
We calculate the energy density matrix and its spectrum for a series of finite systems with 14, 38, 
54, 66, 114, 162, and 186 electrons in cubic periodic cells 
$(L^3=V)$.  These electron counts correspond to closed shell fillings with inversion symmetry in 
momentum space.  The electron density is determined by the parameter 
$r_s\equiv (3V/4\pi N)^{1/3}=3.0$, which is in the metallic regime of the HEG. 

\begin{figure}
\begin{center}
\includegraphics[trim = 0mm 0mm 0mm 0mm, clip,width=1.0\columnwidth]{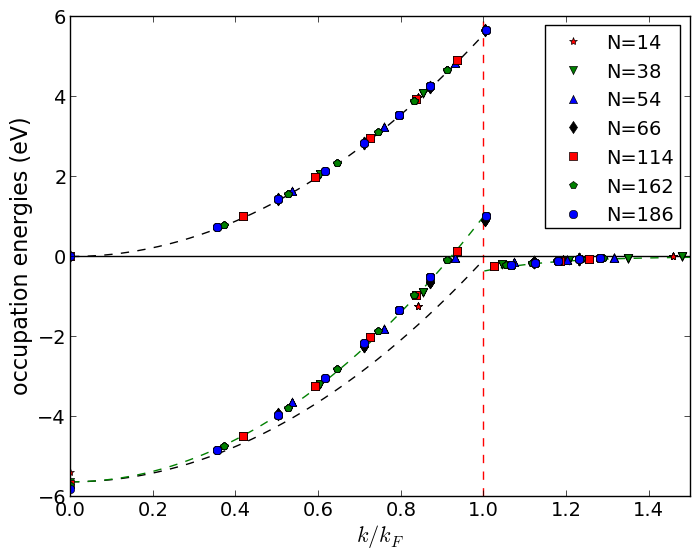}
\end{center}
\caption{Occupation energies for the 3D homogeneous electron gas at $r_s=3$ from diffusion Monte Carlo calculations vs. crystal momentum (solid shapes).  The top panel contains results for the non-interacting system with the black dashed line representing the expected dispersion of $\hbar^2k^2/2m_e$.  Results for the interacting system are in the lower panel.  The dashed green line is a $\hbar^2k^2/2m^*$ fit with $m^*=0.84m_e$ and the black dashed line is the standard dispersion shifted to match at $k=0$.  $N$ is the number of electrons in each finite simulation.\label{fig:heg_spect}}
\end{figure}

The ground state of the non-interacting system is a Slater determinant of plane waves, $e^{ikr}$,  
with each 3-dimensional component $k_d$ of $k$ satisfying $k_d=2\pi n_d/L$.  In this case, the 
exact occupation spectrum must follow the dispersion relation $\mathcal{E}_k=\hbar^2k^2/2m_e$ below 
the Fermi momentum ($k<k_F$) and vanish above it since the unoccupied states do not contribute 
to the total energy.  Since the occupation energies are known for the non-interacting case, 
it serves as a useful test of our implementation of energy density matrix estimators in QMCPACK 
(see appendix \ref{app:qmc_eval} for implementation details).  The occupation energies 
obtained from diffusion Monte Carlo\cite{grimm71,anderson75} (DMC) calculations are shown in the 
top panel of figure 
\ref{fig:heg_spect}.  The solid shapes correspond to energy density matrix (1REDM) eigenvalues, 
$\mathcal{E}_k$, for the various finite systems.  The dashed black curve shows the exact 
dispersion for an infinite system which is strictly positive because the Hamiltonian is comprised of 
kinetic energy only.  The energy spectrum follows the expected parabolic dispersion 
curve for all system sizes, confirming that the energy density matrix estimators have been 
implemented correctly in QMCPACK.

In the presence of electron-electron interactions the natural orbitals and natural energy orbitals 
are constrained by symmetry to remain plane waves but the ground state wavefunction and hence 
its 1RDM and 1REDM become more complex.  Fully interacting occupation energies from the 1REDM are 
shown in the lower panel of figure \ref{fig:heg_spect}.
The dominant effect of interactions is a constant energy shift of just under $6~eV$ for the 
strongly occupied states below the Fermi level, arising partly from the uniform positive 
background present in jellium.  The dashed black line in the lower panel of figure 
\ref{fig:heg_spect} is the non-interacting spectrum shifted to match at $k=0$.  Interactions only 
partly average out with the electron-electron repulsion raising the occupation energies at 
larger $k$ by up to $1~eV$.  Since the dispersion remains approximately parabolic, this effect can 
be summed up in terms of an effective mass, $m^*$.  A dispersion with $m^*=0.84 m_e$ provides a good 
fit to the interacting occupation energies as shown by the dashed green curve.
This suggests a possible relationship between quasiparticle energies and the occupation energies 
since the quasiparticles are known to acquire an effective mass.  Previous quasiparticle calculations 
based on self-consistent $GW$\cite{rietschel83} and related approaches\cite{krakovsky96} find an 
effective mass near $0.9 m_e$.  Effective masses for the two dimensional HEG obtained with  
Quantum Monte Carlo total energy differences are also generally smaller than $GW$ 
results.\cite{kwon94,kwon96,drummond13}

\begin{figure}
\begin{center}
\includegraphics[trim = 0mm 0mm 0mm 0mm, clip,width=1.0\columnwidth]{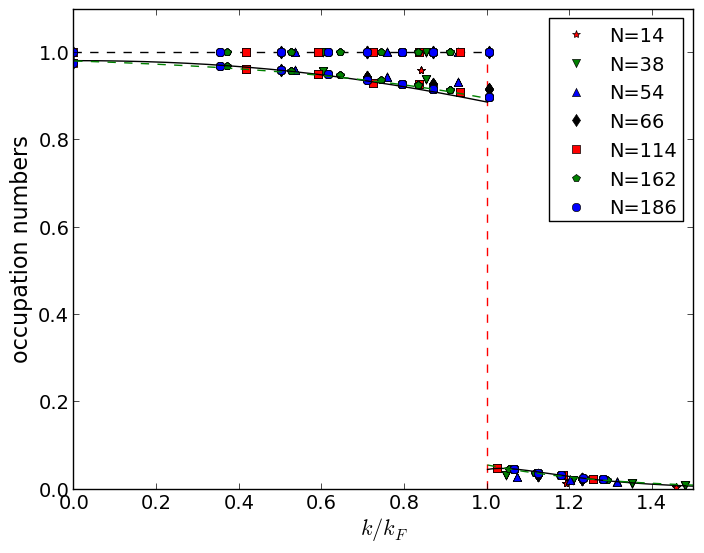}
\end{center}
\caption{Similar to figure \ref{fig:heg_spect} but for occupation numbers of the 1RDM.
         Simulations of interacting systems deviate from one(zero) below(above) $k_F$.  
         The dashed green lines are exponential fits of the deviation from ideal behavior 
         about $k_F$.  For comparison, the solid black curves are polynomial fits to the 
         DMC data of Ortiz and Ballone\cite{ortiz94,ortiz97}. \label{fig:heg_occs}}
\end{figure}

Another known effect of interactions is that the occupation numbers of the 1RDM become finite 
above $k_F$ while the discontinuity at the Fermi surface is reduced.  Figure \ref{fig:heg_occs}
shows the DMC eigenvalues of the 1RDM vs. particle count from our calculations.  The occupation 
numbers above the Fermi surface decrease exponentially with increasing momentum.  Despite the 
rapid decay, fully $7\%$ of the particle weight resides above the Fermi surface.  
This effect is also visible in the occupation energy spectrum above $k_F$ in figure \ref{fig:heg_spect}.  
Similar to the occupation numbers, the occupation energies also decrease exponentially with $k$ 
above the Fermi surface and cumulatively account for $10\%$ of the total energy.  
We expect this behavior to generalize to other systems with strongly 
occupied states below the Fermi level (which carry most of the energy) being accompanied by a 
large number of weakly occupied states providing evanescent energy contributions. 
The fraction of the total energy residing above the Fermi level, as obtained from the 
energy density matrix, can be viewed as a measure of correlation.  With this metric, 
the correlation energy of any mean field system is precisely zero.

\subsection{Occupation energies for selected excitations of a 14 electron HEG}
\label{sec:esheg}
In order to explore the relationship between the occupation energies obtained from the 1REDM and 
electron addition and removal energies, we have performed additional DMC calculations on the 14 
electron system.  The first of these is a direct charge removal from the outermost filled spin 
down shell performed at constant volume.  The second is a spin flip excitation where an electron 
in the outermost spin down shell is promoted to a new unfilled spin up shell.  In both cases we 
compare total energy differences relative to the unperturbed 14 electron state with results from 
the active occupation energies.  
The active space is defined as the set of natural energy orbitals 
that experienced a significant change in the spin-resolved occupation number between the initial and 
final states.  Eigenstates that did not significantly 
vary in occupation number are further separated into core states (with occupation number closer to one) 
and virtual states (with occupation number closer to zero).
Although the comparison is made here for a small system for simplicity and convenience, the 
conclusions drawn from these results should not qualitatively depend on system size.

\begin{figure}
\begin{center}
\includegraphics[trim = 0mm 0mm 0mm 0mm, clip,width=1.0\columnwidth]{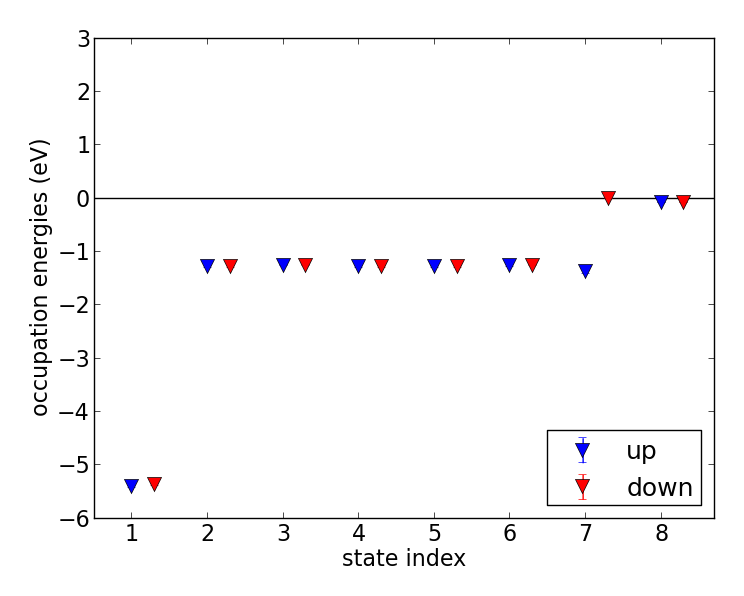}
\end{center}
\caption{DMC occupation energies for a 13 electron HEG vs. state index (solid shapes) resolved by spin with spin up in blue and spin down in red.  State 1 corresponds to $k=0$ and states 2 through 7 belong to the same shell $k\in \{\pm \Delta k \hat{x},\pm \Delta k \hat{y},\pm \Delta k \hat{z}\}$.  A single electron has been removed from the spin down channel of state 7.\label{fig:heg_charge_rem}}
\end{figure}

Spin resolved occupation energies (see Eq. \ref{eq:Eocc_spin}) for the system with a single 
electron removed are shown in figure \ref{fig:heg_charge_rem}.  Spin up(down) states are 
represented by the blue(red) triangles.  The $k=0$ state (state 1) is lowest in energy and 
all 6 states in the second shell ($\abs{k}=\Delta k$ are degenerate.  The electron removal 
is clearly visible with the previously degenerate 7th spin down state displaying an 
occupation energy near zero.  
Only this state (state 7) has experienced a significant change in occupation number 
and so the active space contains this state alone.  Since the neighboring spectrum is changed 
only a small amount, total energy differences will be dominated by the active space as core 
and virtual contributions to energy differences largely cancel.

Charged systems in periodic boundary 
conditions experience a shift in the reference potential relative to the neutral state\cite{corsetti2011,komsa2012}.  
The relative difference in occupation energy of the first and second shells for the neutral 
and charged systems is identical to within error bars and so the potential shift can safely 
be obtained by aligning the core levels.  This is effected by applying a constant potential 
energy shift of $-0.11~eV$ to the charged system.  Calculating the potential shift from 
core occupation energies might also prove useful to QMC studies of charged defect systems 
which are generally restricted to small supercells.

\begin{figure}
\begin{center}
\includegraphics[trim = 0mm 0mm 0mm 0mm, clip,width=1.0\columnwidth]{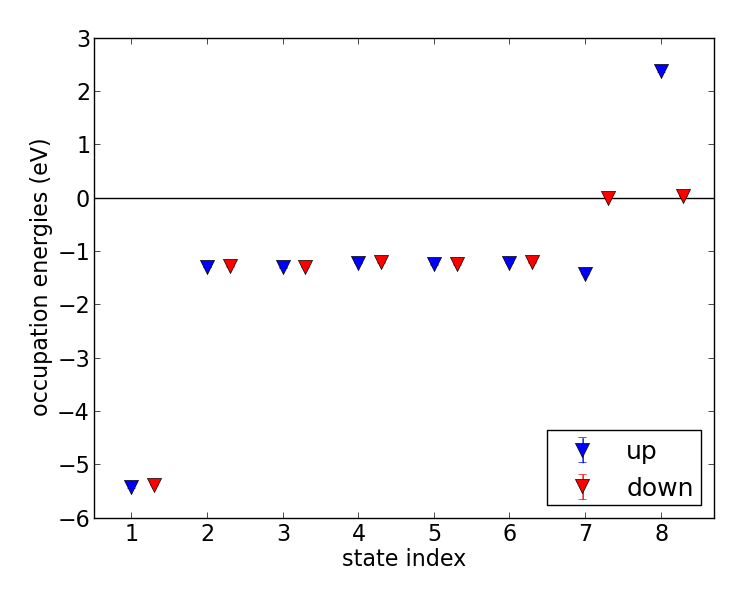}
\end{center}
\caption{DMC occupation energies for a 14 electron HEG spin flip excitation vs. state index (solid shapes) resolved by spin with spin up in blue and spin down in red.  A single electron has been removed from the spin down channel of state 7 and added to the spin up channel of state 8, which belongs to the third shell, $k=2\Delta k \hat{x}$.\label{fig:heg_spin_flip}}
\end{figure}

Having taken the alignment potential into account, a meaningful comparison of total energy 
differences and changes in the active occupation energies can be made.  The charge removal 
energy obtained from DMC total energy differences is 
$\Delta E_{tot}=E_{tot}^{7^\uparrow6^\downarrow}-E_{tot}^{7^\uparrow7^\downarrow}=1.43(2)~eV$.  For 
comparison the change in occupation energy over the active space is 
$\Delta E_{active}=\mathcal{E}_{7^\downarrow}^{7^\uparrow6^\downarrow}-\mathcal{E}_{7^\downarrow}^{7^\uparrow7^\downarrow}=1.27(1)~eV$.
The occupation energy of the active state in the neutral system accounts for about $90\%$ 
of the ionization energy ($-\mathcal{E}_{7^\downarrow}^{7^\uparrow7^\downarrow}=1.28(1)~eV$).  The other $10\%$ 
is scattered across the core and virtual spaces as the result of correlation, which is consistent 
with the fraction of total energy residing above the Fermi level we have already witnessed in the 
ground state.  Although some of the energy is dispersed across the 
state space, the large concentration of energy residing in a single state indicates that 
the occupation energies of the 1REDM are closely related to particle addition and removal 
energies for this system.

This relationship is also confirmed in the case of a spin flip excitation.  Occupation 
energies for a 14 electron HEG with 8 up and 6 down electrons can be found in figure 
\ref{fig:heg_spin_flip}.  The spectrum is similar to the charge removal case for states 
7 and below since the spin flip consists of removing an electron from the spin down 
channel of state 7 and then adding it to the spin up channel of state 8, which resides 
on a higher energy shell.  The active space is comprised of these two states.  
Since the filled shell and spin-flipped systems have the same charge the constant 
background potentials are already aligned and the spectra can be compared directly.  
The energy required to flip the spin is 
$\Delta E_{tot}=E_{tot}^{8^\uparrow6^\downarrow}-E_{tot}^{7^\uparrow7^\downarrow}=3.98(2)~eV$ 
according to DMC total energies and 
$\Delta E_{active}=\mathcal{E}_{8^\uparrow}^{8^\uparrow6^\downarrow}+\mathcal{E}_{7^\downarrow}^{8^\uparrow6^\downarrow}-\mathcal{E}_{8^\uparrow}^{7^\uparrow7^\downarrow}-\mathcal{E}_{7^\downarrow}^{7^\uparrow7^\downarrow}=3.71(3)~eV$ 
according to the active occupation energies.  The fraction of energy contained in the 
active space for the spin flip is similar to the charge removal case.

\subsection{Occupation energies of the oxygen atom}
\label{sec:gsox}

\begin{figure}
\begin{center}
\includegraphics[trim = 0mm 0mm 0mm 0mm, clip,width=1.0\columnwidth]{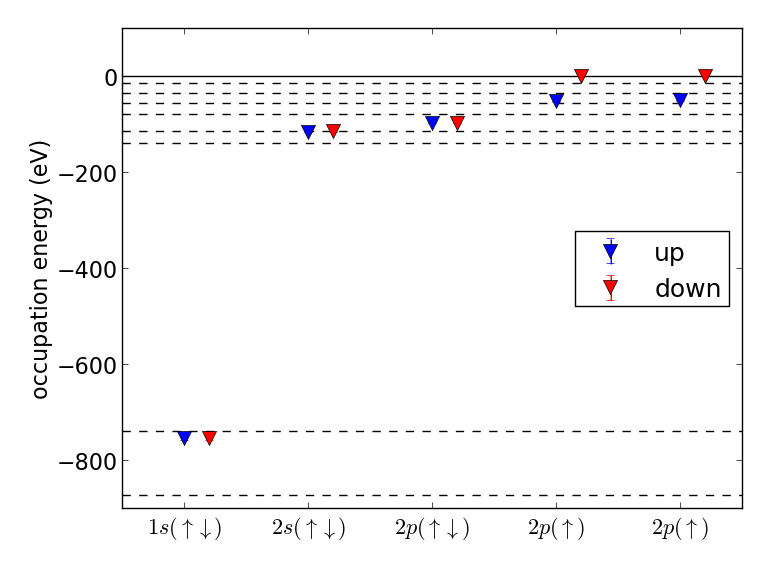}
\end{center}
\caption{DMC occupation energies for the ground state of a neutral oxygen atom 
with 5 up and 3 down electrons (solid triangles).   Experimental binding energies 
(negative of ionization potential) of the first through the eighth electron in 
oxygen are shown as dashed horizontal lines as a comparative energy scale.\label{fig:aeox}}
\end{figure}

Diffusion Monte Carlo occupation energies obtained with QMCPACK for an all-electron 
oxygen atom in its electronic ground state are shown in figure \ref{fig:aeox}.  In 
this case, the energy density matrix has been expanded in a basis of 46 orbitals 
obtained from a Hartree-Fock\cite{hartree28,fock30,roothaan51} calculation performed 
with the GAMESS\cite{schmidt93,gordon05} simulation package.  
The spin state of the atom is clearly reflected in the occupation spectrum with three 
electron pairs occupying $1s$, $2s$, and $2p$ states along with two unpaired spin down 
(blue triangles) electrons  in the outermost valence states.  The spin up 
(red triangles) partner orbitals of the unpaired electrons are unoccupied as 
indicated by their near zero occupation energies.  The negatives of experimental 
ionization energies (\emph{i.e.} the binding energies of single electrons for 
successive ionizations) are shown in dashed black lines to give a comparative energy 
scale.  This comparison is useful because these binding energies and the occupation 
energies must add up to the total energy ($E_{tot}^{exp}=-2043.81~eV$, 
$E_{tot}^{DMC}=-2042.29(4)~eV$).  The $1s$ electrons are very deeply 
bound in the ground state with occupation energies of $-753(4)~eV$.  This value 
resides fairly close to the experimental binding energy of $-739.29~eV$ for the 
first $1s$ electron of the $O^{6+}$ ion.  This demonstrates that the occupation spectrum 
has remained bounded in a physical way despite the strong inhomogeneity of the nucleus.  

Moving up the spectrum, we see that the ground state $2s$ electron occupation 
energies are similarly close to the experimental binding energy of the first $2s$ 
electron of the $O^{4+}$ ion and so the large gap between $1s$ and $2s$ states is 
represented well.  Out in the $2p$ valence states we see a rather different distribution 
of energy in the ground state relative to energies imparted to electrons upon ionization. 
The occupation energies of the $2p$ states are rather deep.  This implies that 
electronic relaxation effects will be quite pronounced for the occupation energies 
of the 1REDM, with many occupation energies changing during ionization.  
It is not fully clear why the occupation energies resemble electron removal 
energies rather well for the homogeneous electron gas and poorly for inhomogeneous 
oxygen.  It is true that the occupation energies resulting from the energy density 
matrix are not formally guaranteed to obey Koopman's theorem except in the 
weakly-interacting limit.  In this respect the correspondence between the occupation 
energies and particle addition/removal energies will depend on the system being 
studied.
The occupation 
energies in figure \ref{fig:aeox} are on the same scale as the experimental binding 
energies and so they do remain related, although not as closely, to electron removal 
energies for this system.

\section{Summary} \label{sec:summary}
We have introduced a new observable for many-body quantum systems, the energy 
density matrix, analogous to the well established one body reduced density 
matrix.  The natural orbitals of the two matrices are similar, with the 1RDM 
providing particle number information in the form of occupation numbers and 
the 1REDM providing a complementary description of energetics in the form 
of occupation energies.  
We have also argued that the evanescent portion of the occupation energy 
spectrum and the deviations between the two sets of orbitals are a signal 
of strong correlation effects.
It has been shown that the occupation energies 
obey a total energy sum rule, similar to both the mean field case and 
the Landau total energy functional for quantum fluids.  We have also 
demonstrated that the eigenstates of the energy density matrix obey a 
Schr\"{o}dinger-like equation that is similar in form to the standard 
quasiparticle equation.  The resulting occupation energies for the homogeneous 
electron gas in the metallic regime resemble quasiparticle energies, in the 
sense that the spectrum can be described in terms of an effective mass.  The 
occupation energies also approximate electron addition and removal energies 
for this system as demonstrated by direct comparison with total energy 
differences for simple charge and spin excitations.  In the inhomogeneous 
case, studied here for a single oxygen atom, the occupation spectrum of 
the ground state remains bounded from below and reproduces some features 
of the experimental electron binding energies.  The overall quantitative 
agreement is rather poorer than for HEG, however, 
showing that the correspondence is not fully general. 
With further development the energy density 
matrix may provide a useful tool to describe single particle energetics 
with quantum Monte Carlo methods which have traditionally been limited 
to total energies.

\section*{Acknowledgements}
The authors (JTK, JK, \& FR) would like to thank Paul Kent for a thorough reading 
of the manuscript and useful discussions during the development of this study. 
The work was supported by the Materials Sciences \& Engineering Division 
of the Office of Basic Energy Sciences, U.S. Department of Energy.  One of us 
(JK) was supported through the Predictive Theory and Modeling for Materials and 
Chemical Science program by the Basic Energy Science (BES), Department of Energy (DOE).

\appendix
\section{Quantum Monte Carlo evaluation} \label{app:qmc_eval}

This section details the practical evaluation of the energy density matrix within 
standard ground-state continuum quantum Monte Carlo methods such as variational\cite{mcmillan65} 
(VMC) or diffusion Monte Carlo (DMC).  The formal details and recent applications 
of these methods have been covered 
elsewhere\cite{foulkes01,bajdich09,needs10,kolorenc11,wagner14} 
and we refer the interested reader to these sources 
to obtain a full account.  These methods are treated here in the abstract, but 
sufficient detail is retained to unambigously sample the energy density matrix 
in a real simulation code.

In DMC one measures observables relative to the mixed $N$-body density matrix
\begin{align}
  \hat{\rho}^{DMC}_N = \lvert \Psi \rangle\langle \Psi_0 \rvert.
\end{align}
Here $\Psi$ represents the analytically defined trial wavefunction and $\Psi_0$ 
is the fixed node/phase approximation to the ground state as produced by the 
diffusion and branching process of Monte Carlo walkers.  VMC results are 
obtained by the substitution $\Psi_0\rightarrow \Psi$.  The expectation value of 
observable $A$ is obtained in the usual way:
\begin{align}
  \mean{A} &= Tr \hat{A}\hat{\rho}^{DMC}_N = \expval{\Psi_0}{A}{\Psi} \nonumber\\
           &= \int dR \Psi^*_0(R)\Psi(R) \frac{A\Psi(R)}{\Psi(R)} .
\end{align} 
The DMC process explicitly draws configuration space samples $\{R\}$ from the 
mixed probability distribution $\Psi^*_0(R)\Psi(R)$.  

An efficient and compact representation of the energy density matrix can be 
obtained by projection onto a suitably chosen single particle basis $\{\phi_i\}$.
In the case of the HEG these are plane waves, while for systems composed of 
atoms it is convenient to use orbitals from Hartree-Fock or DFT calculations.  
In the low energy subset of the basis, a finite and discrete approximation to the
energy density matrix ($\mathcal{E}_{1ij}$) is obtained.
\begin{align}\label{eq:qmc_naive}
  \mathcal{E}_{1ij} &= \expvalnh{\phi_i}{\hat{\mathcal{E}}_1}{\phi_j} \nonumber \\
                   &= \expvalnh{\phi_i}{\sum_nTr_{R_n}\hat{h}_{n}\hat{\rho}^{DMC}_N}{\phi_j} \nonumber \\
                   &= \sum_n \int dR_ndr_ndr'_n \phi_i^*(r_n') h(r_n',R_n) \nonumber\\
                   &  \qquad \qquad \times \Psi(r_n',R_n)\Psi_0^*(r_n,R_n)\phi_j(r_n) \nonumber \\
                   &=  \int dR \Psi_0^*(R)\Psi(R) \sum_n\phi_j(r_n)  \nonumber \\
                   &  \times \int dr'_n \frac{\Psi(r_n',R_n)}{\Psi(r_n,R_n)}\phi_i^*(r_n')  \frac{h(r_n',R_n)\Psi(r_n',R_n)}{\Psi(r_n',R_n)}  
\end{align}
For systems involving non-local pseudopotentials, the factor $h(r_n',R_n)\Psi(r_n',R_n)$ 
is replaced with $\int d\bar{r}_nh(r_n',\bar{r}_n,R_n)\Psi(\bar{r}_n,R_n)$, where $\bar{r}_n$ 
represents the additional non-local coordinate.

Equation \ref{eq:qmc_naive} is a valid way to measure $\hat{\mathcal{E}}_1$, but it 
is inefficient since the additional integral over $r_n'$ involves a re-evaluation 
of the Hamiltonian components $h(r_n',R_n)$ at each integration point.
A more efficient form can be obtained without a loss of accuracy.  Upon switching 
the primed coordinates $r_n\leftrightarrow r_n'$ and rearranging, we obtain 
\begin{align}\label{eq:qmc_fast}
  \mathcal{E}_{1ij} &=  \sum_n \int dR dr_n' \Psi^*(R)\Psi(R) \frac{\Psi_0^*(r_n',R_n)}{\Psi(r_n,R_n)}\phi_i^*(r_n) \nonumber\\
                   & \quad \times \phi_j(r_n') \frac{h(r_n,R_n)\Psi(R)}{\Psi(R)}  \nonumber \\
                   &\approx  \sum_n \int dR dr_n' \Psi_0^*(R)\Psi(R) \frac{\Psi^*(r_n',R_n)}{\Psi^*(r_n,R_n)}\phi_i^*(r_n) \nonumber\\ 
                   & \quad \times \phi_j(r_n') \frac{h(r_n,R_n)\Psi(R)}{\Psi(R)}  \nonumber \\
                   &=   \int dR  \Psi_0^*(R)\Psi(R) \sum_n \frac{h(r_n,R_n)\Psi(R)}{\Psi(R)} \phi_i^*(r_n)  \nonumber \\
                   & \quad \times \int dr_n' \frac{\Psi^*(r_n',R_n)}{\Psi^*(r_n,R_n)} \phi_j(r_n')   
\end{align}
This representation is more efficient because the quantities involving $h(r_n,R_n)$
have already been computed at each Monte Carlo configuration $R$.  In this way 
the energy density matrix can be computed at no additional cost over the 1RDM.
The additional integral over $r_n'$ can be evaluated approximately as a Riemann 
sum over a randomly shifted uniform grid, as is done in this work, 
or from a set of points sampled from 
the single particle density.  Reusing the same set of points for each value 
of $n$ additionally reduces the number of required orbital evaluations by 
a factor of $N$, but the relatively expensive wavefunction ratios must still 
be computed for each $n$.

We will now show that the approximation in Eq. \ref{eq:qmc_fast} does not affect 
the accuracy of the computed energy density matrix.  For simplicity of discussion we will 
actually consider the 1RDM since the approximation above affects the sampling of 
each matrix in the same fashion.  The 1RDM in DMC is
\begin{align}
  n_1(r,r') = \sum_n \int dR_n \Psi(r,R_n)\Psi_0^*(r',R_n).
\end{align}
Considering instead the 1RDM arising from the adjoint of $\hat{\rho}^{DMC}_N$
\begin{align}
    n_1(r,r')^\dagger &= \sum_n \int dR_n \Psi_0(r,R_n)\Psi^*(r',R_n) \nonumber \\
                   &= \sum_n \int dR_n \Psi(r,R_n)\Psi_0^*(r',R_n) \nonumber \\
                   & \qquad \times \frac{\Psi_0(r,R_n)}{\Psi(r,R_n)}\frac{\Psi^*(r',R_n)}{\Psi_0^*(r',R_n)}
\end{align}
we see that it differs from the original by the kernel 
\begin{align}
  K^T(r,r',R_n) =  \frac{\Psi_0(r,R_n)}{\Psi(r,R_n)}\frac{\Psi^*(r',R_n)}{\Psi_0^*(r',R_n)}.
\end{align}
The approximation in equation \ref{eq:qmc_fast} is equivalent to introducing 
the kernel 
\begin{align}
  K^A(r,r',R_n) =  \frac{\Psi_0^*(r,R_n)}{\Psi^*(r,R_n)}\frac{\Psi^*(r',R_n)}{\Psi_0^*(r',R_n)}.
\end{align}
For VMC $\Psi_0\rightarrow \Psi$ and $K^A_{VMC}=K^T_{VMC}=1$.  For fixed-node (FN) and released-node (RN)
 DMC the wavefunction is real ($\Psi_0^*=\Psi_0$, $\Psi^*=\Psi$) and 
\begin{align}
  K^A_{FN}=K^T_{FN}=\frac{\Psi_0(r,R_n)}{\Psi(r,R_n)}\frac{\Psi(r',R_n)}{\Psi_0(r',R_n)}.  
\end{align}
Fixed-phase (FP) DMC gives a similar result since the trial wavefunction and its projection are 
contrained to share the same phase ($\Psi_0=\abs{\Psi_0}e^{i\phi}$, $\Psi=\abs{\Psi}e^{i\phi}$) 
which yields 
\begin{align}
  K^A_{FP}=K^T_{FP}=\frac{\abs{\Psi_0(r,R_n)}}{\abs{\Psi(r,R_n)}}\frac{\abs{\Psi(r',R_n)}}{\abs{\Psi_0(r',R_n)}}.
\end{align} 
In all of these cases the 1RDM and the 1REDM are effectively being measured from $\hat{\rho}_N^\dagger$ 
which has the same level of accuracy as  $\hat{\rho}_N$.  Additionally, the sum rules summarized in 
table \ref{tab:comp} are all preserved since $K^A(r,r,R_n)=1$.

\bibliography{ref}

\end{document}